\documentclass[nofootinbib, preprint]{revtex4} 
\usepackage{amssymb, amsmath}
\usepackage{color, soul}
\usepackage{graphicx}
\usepackage{wrapfig}
\usepackage{epstopdf}
\begin{document}
\title{A local $\psi$-epistemic retrocausal hidden-variable model of Bell correlations with wavefunctions in physical space}
\author{Indrajit Sen}
\affiliation{Department of Physics and Astronomy,
Clemson University, Kinard Laboratory,
Clemson, SC 29634, USA}
\date{\today}
\email{isen@g.clemson.edu}

\begin{abstract}
We construct a local $\psi$-epistemic hidden-variable model of Bell correlations by a retrocausal adaptation of  the originally superdeterministic model given by Brans. In our model, for a pair of particles the joint quantum state  $|\psi_e(t)\rangle$ as determined by preparation is epistemic. The model also assigns to the pair of particles a factorisable joint quantum state $|\psi_o(t)\rangle$ which is different from the prepared quantum state $|\psi_e(t)\rangle$ and has an ontic status. The ontic state of a single particle consists of two parts. First, a single particle ontic quantum state $\chi(\vec{x},t)|i\rangle$, where $\chi(\vec{x},t)$ is a 3-space wavepacket and $|i\rangle$ is a spin eigenstate of the \textit{future} measurement setting. Second, a particle position in 3-space $\vec{x}(t)$, which evolves via a de Broglie-Bohm type guidance equation with the 3-space wavepacket $\chi(\vec{x},t)$ acting as a local pilot wave. The joint ontic quantum state $|\psi_o(t)\rangle$ fixes the measurement outcomes deterministically whereas the prepared quantum state $|\psi_e(t)\rangle$ determines the distribution of the $|\psi_o(t)\rangle$'s over an ensemble. Both $|\psi_o(t)\rangle$ and $|\psi_e(t)\rangle$ evolve via the Schrodinger equation. Our model exactly reproduces the Bell correlations for any pair of measurement settings. We also consider `non-equilibrium' extensions of the model with an arbitrary distribution of hidden variables. We show that, in non-equilibrium, the model generally violates no-signalling constraints while remaining local with respect to both ontology and interaction between particles. We argue that our model shares some structural similarities with the modal class of interpretations of quantum mechanics.
\end{abstract}

\maketitle

\section{Introduction}\label{sec1}
One of the most important contributions to the long standing debate about the physical interpretation of Quantum Mechanics (QM) is Bell's theorem \cite{bell}, which proved that any hidden-variable completion of QM as envisaged by EPR \cite{epr} must be nonlocal. It has however often been under-emphasized in the literature that the theorem makes an important assumption about the relationship between the hidden-variables and the measurement settings.\\

The assumption is that the hidden variables describing the quantum systems, and the measurements that these systems are subjected to in future, are uncorrelated. That is, the following is assumed about the hidden-variable distribution:
\begin{align}
\rho(\lambda||\psi\rangle, M) = \rho(\lambda||\psi\rangle)
\end{align}
where the hidden-variables are labelled by $\lambda$, the preparation by quantum state $|\psi\rangle$, and the observable being measured (or the measurement basis) by $M$. This assumption, often termed as Measurement Independence \cite{hall10, howmuch} in recent literature, is necessary to rule out local hidden-variable models of QM via Bell's theorem. There are atleast two physically different kinds of hidden-variable models where Measurement Independence fails, thereby circumventing the theorem. \\

\textit{Superdeterministic} models posit that the hidden variables and the measurement settings are correlated by common causes in the past. Such models attempt to explain the Bell correlations by yet another correlation, now at the hidden-variable level - correlation between the hidden variables which describe the quantum systems and the hidden variables which determine the measurement settings, due to past common causes. However, how can we be sure such common causes \textit{always} exist whenever a Bell inequality violation is observed, or that such correlations at the hidden-variable level are \textit{exactly} of the magnitude to reproduce the Bell correlations at the quantum level, each time? For such reasons they have been widely criticised in the literature as `conspiratorial' \cite{pricebash, pironio}, with some important exceptions \cite{brans, hooft}. Recently experiments, which employ cosmic photons to determine the measurement settings, have been proposed \cite{cosmicbellI} and conducted \cite{cosmicbellII} which severely constrain these models.\\

\textit{Retrocausal} models on the other hand posit that the measurement settings act as a cause (in the future) to affect the hidden-variable distribution during preparation (in the past). This is highly counterintuitive to our sense of causality and time, but its proponents \cite{cramer, pricebook, tsvfupdate} claim it is our latter notions that are suspect at the microscopic level. Both kinds of models have implications for the most important questions in the interpretation of QM - the reality of the quantum state, and nonlocality, but there are at present few models of either type in literature. In this article we present a local retrocausal model of Bell correlations, adapting a model given by Brans \cite{brans} in the 1980s, who presented it as an example to argue in favour of superdeterminism. The Brans model itself has been generalised to arbitrary preparations and measurements \cite{hall16}, and proven to be maximally $\psi$-epistemic in any number of dimensions of Hilbert space \cite{mythesis}. \\

We also consider arbitrary distributions of hidden variables in our model, that do not reproduce the Bell correlations. Valentini \cite{valentini, genon} has argued that hidden-variable models must accomodate non-fine-tuned or `non-equilibrium' distributions which do not reproduce the QM predictions. This is because initial conditions do not have the status of a law in a theory, but are instead contingent. The same conclusion can also be drawn from the more recent work by Wood and Spekkens \cite{woodspek}, who have criticised causal explanations of Bell correlations as being `conspiratorial', in the sense that such models require a fine tuning in the hidden-variable distribution to be non-signalling. If we take the concept of a hidden variable model underlying QM seriously, it follows that QM is a special case of a fine-tuned distribution in the hidden-variable model, which itself contains a much wider physics described by non-equilibrium distributions. Astrophysical and cosmological tests for the existence of such non-equilibrium distributions have been proposed \cite{valentiniastro}. We therefore discuss a non-equilibrium extension of our model, and explore the interplay between locality, retrocausality and no-signalling.\\

The structure of this paper is as follows. We first take the superdeterministic model given by Brans and present its equations without invoking any physical interpretation (superdeterministic or retrocausal or otherwise) of correlation between the hidden variables and the measurement settings. Then we provide a retrocausal interpretation, present our model in detail and show how it reproduces the Bell correlations. Next we consider non-equilibrium extensions of the model, and show that a non-fine-tuned distribution of hidden variables leads to nonlocal signalling in general. We conclude by discussing properties of the model and its connection with modal interpretations of QM.

\section{The Brans model}
Consider the standard Bell scenario \cite{bell1975}, where two spin-1/2 particles are prepared in a spin-singlet state and then local measurements $\hat{\sigma}_{\hat{a}}\otimes \hat{I}$ and $\hat{I} \otimes \hat{\sigma}_{\hat{b}}$ are subsequently performed on the particles at a spacelike separation\footnote{$\hat{\sigma}_{\hat{a}(\hat{b})}\equiv\hat{\sigma}\cdot\hat{a}(\hat{b})$}. Let $\lambda'_i$ and $\lambda'_j$, $i,j \in \{+,-\}$, be local hidden-variables describing the two particles, and let their distribution be given by
\begin{align}
&p(\lambda'_i, \lambda'_j| |\psi_{singlet}\rangle, \hat{\sigma}_{\hat{a}}, \hat{\sigma}_{\hat{b}}) = |\langle \psi_{singlet}|(|i\rangle_{\hat{a}}\otimes  |j\rangle _{\hat{b}}) |^2 \label{o}
\end{align}
where $|i\rangle_{\hat{a} (\hat{b})}$ denotes an eigenstate of $\hat{\sigma}_{\hat{a}(\hat{b})}$.
The local outcomes are specified by \\
\begin{align}
& A(\lambda'_i) = i \label{a'}\\
& B(\lambda'_j) = j \label{b'}
\end{align}
The model reproduces Bell correlations:

\begin{align}
\langle\hat{\sigma}_{\hat{a}} \otimes\hat{\sigma}_{\hat{b}}\rangle &= \sum_{ij} A(\lambda'_i) B(\lambda'_j) p(\lambda'_i, \lambda'_j | |\psi_{singlet}\rangle, 
\hat{\sigma}_{\hat{a}}, \hat{\sigma}_{\hat{b}}) \\
& = \sum_{ij} i.j. |\langle \psi_{singlet}|( |i\rangle_{\hat{a}}\otimes  |j\rangle _{\hat{b}}) |^2  \\
& = |\langle \psi_{singlet}|( |+\rangle_{\hat{a}}\otimes  |+\rangle _{\hat{b}}) |^2 - |\langle \psi_{singlet}|( |+\rangle_{\hat{a}}\otimes  |-\rangle _{\hat{b}}) |^2 \nonumber \\
& - |\langle \psi_{singlet}|( |-\rangle_{\hat{a}}\otimes  |+\rangle _{\hat{b}}) |^2 
+ |\langle \psi_{singlet}|( |-\rangle_{\hat{a}}\otimes  |-\rangle _{\hat{b}}) |^2
\end{align}

The model satisfies locality and determinism, from eqns. \ref{a'} and \ref{b'}. But it does not satisfy Measurement Independence from equation \ref{o}, as the hidden-variable distribution depends on the measurement settings $\hat{\sigma}_{\hat{a}}$ and $\hat{\sigma}_{\hat{b}}$.

\section{A retrocausal interpretation of the Brans model}\label{main}
We now lend a retrocausal interpretation to the equations of the Brans model. We first posit that the information about measurement settings made in the \textit{future}, $\hat{\sigma}_{\hat{a}}\otimes \hat{I}$ and $\hat{I} \otimes \hat{\sigma}_{\hat{b}}$, is made available to the particles at the preparation source in the \textit{past}, by an as yet not understood `retrocausal mechanism'. This causes the particles to be \textit{prepared in one of the eigenstates of the future measurement settings}. . That is, the pairs of particles are prepared in one of these joint spin states: $|+\rangle_{\hat{a}} \otimes|+\rangle _{\hat{b}}$, $|+\rangle_{\hat{a}} \otimes |-\rangle _{\hat{b}}$, $|-\rangle_{\hat{a}}\otimes |+\rangle _{\hat{b}}$, $|-\rangle_{\hat{a}} \otimes|-\rangle _{\hat{b}}$\footnote{The role of the preparation-determined quantum state $|\psi_{singlet}\rangle$ in our model is explained below.}.\\

Hence each particle is described in our model by an \textit{ontic quantum state} of the form $\chi(\vec{r},t)|i\rangle$, $i \in \{+,-\}$, where $\chi(\vec{r},t)$ is a single particle 3-space wavepacket and $|i\rangle$ is an eigenstate of the future measurement setting. The pair of particles is described by the initial joint ontic quantum state $\langle \vec{r}_1| \langle \vec{r}_2|\psi_o(0)\rangle = \chi_1(\vec{r}_1,0)|i_1\rangle_{\hat{a}} \otimes\chi_2(\vec{r}_2,0)|i_2\rangle_{\hat{b}}$. We term the preparation-determined quantum state $\langle \vec{r}_1| \langle \vec{r}_2|\psi_e(0)\rangle=\chi_1(\vec{r}_1,0)\chi_2(\vec{r}_2,0)|\psi_{singlet}\rangle$ as the \textit{epistemic quantum state}. Both the joint ontic quantum state $\big ($with two single particle 3-space wavepackets $\chi_1(\vec{r}_1,t)$ and $\chi_2(\vec{r}_2,t)\big)$ and the epistemic quantum state $\big($with a single configuration space wavepacket $\chi_{12}(\vec{r}_1,\vec{r}_2, t)$ in general$\big)$ evolve via the Schrodinger equation in our model. \\
\begin{figure}
\centering
\includegraphics[scale=0.7]{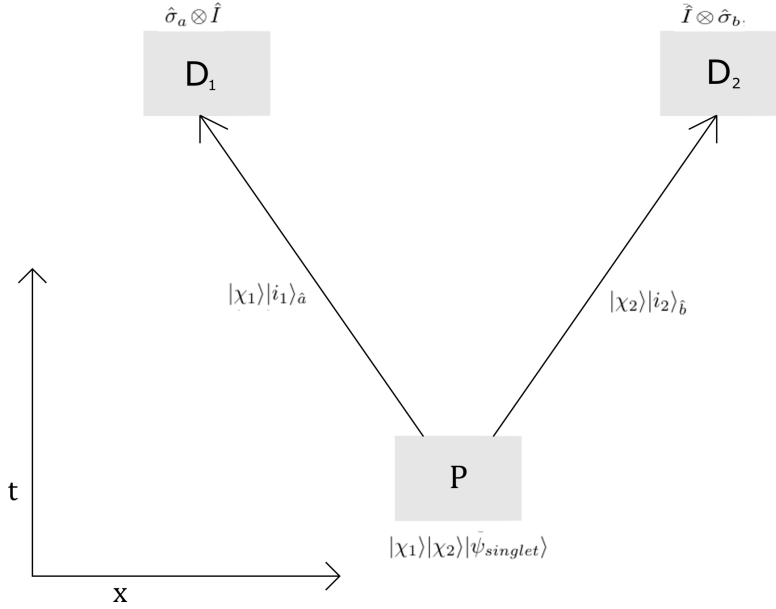}
\caption{Schematic illustration of our model. A preparation device P prepares a quantum state $\langle \vec{r}_1| \langle \vec{r}_2|\psi_e(0)\rangle = \chi_1(\vec{r}_1,0)\chi_2(\vec{r}_2,0)|\psi_{singlet}\rangle$ of the two spin-1/2 particles. The detectors $D_1$ and $D_2$ are set, at the spacetime regions indicated, to measure the observables $\hat{\sigma}_{\hat{a}}\otimes \hat{I}$ and $\hat{I}\otimes\hat{\sigma}_{\hat{b}}$ respectively. This information about the measurement settings is made available at P retrocausally, and fixes the ontic quantum state of the two particles to an eigenstate $\langle \vec{r}_1| \langle \vec{r}_2|\psi_o(0)\rangle=\chi_1(\vec{r}_1,0)|i_1\rangle_{\hat{a}}\otimes\chi_2(\vec{r}_2,0)|i_2\rangle_{\hat{b}}$, where $i_1,i_2 \in \{+,-\}$ are chosen randomly. Both $|\psi_e(t)\rangle$ and $|\psi_o(t)\rangle$ evolve via the Schrodinger equation. The wavepackets $\chi_1(\vec{r}_1,t)$ and $\chi_2(\vec{r}_2,t)$ act as local pilot waves for the corresponding particles via equation \ref{d4}. The resulting dynamics deterministically fixes the measurement outcomes for an individual case. The preparation-determined quantum state $|\psi_e(t)\rangle$ plays a purely statistical role of determining the distribution of the various ontic quantum states for an ensemble.}
\end{figure}

We next posit that each particle has a definite position at all times, with velocity given by 
\begin{align}
\vec{v} = \frac{\vec{\nabla}S(\vec{r},t)}{m}\label{d4}
\end{align}
where $\chi(\vec{r},t)=R(\vec{r},t)e^{iS(\vec{r},t)}$ is the 3-space wavepacket of that particle, contained in the ontic quantum state. The trajectory of the particle (and hence the measurement outcome) is thus determined locally by the single-particle ontic quantum state. This completes description of the ontology of our model. We now turn to describe the distribution of these hidden variables for an ensemble of pairs of particles having the same epistemic quantum state $|\psi_e(0)\rangle$.\\

We first assume that the expansion of the preparation-determined (epistemic) quantum state in the future measurement basis 
\begin{align}
|\psi_{singlet}\rangle = c_{++}|+\rangle _{\hat{b}} + c_{+-}|-\rangle _{\hat{b}}
+ c_{-+} |+\rangle _{\hat{b}} + c_{--} |-\rangle _{\hat{b}} \label{d3}
\end{align}
determines the ensemble-proportions $|c_{++}|^2$, $|c_{+-}|^2$, $|c_{-+}|^2$, $|c_{--}|^2$ of the initial joint ontic quantum states $\chi_1(\vec{r}_1,0)|+\rangle_{\hat{a}}\otimes \chi_2(\vec{r}_2,0) |+\rangle _{\hat{b}}, \textbf{ }\chi_1(\vec{r}_1,0)|+\rangle_{\hat{a}}\otimes \chi_2(\vec{r}_2,0)|-\rangle _{\hat{b}}, \text{ }\chi_1(\vec{r}_1,0)|-\rangle_{\hat{a}} \otimes \chi_2(\vec{r}_2,0) |+\rangle _{\hat{b}}, \text{ }\chi_1(\vec{r}_1,0)|-\rangle_{\hat{a}} \otimes \chi_2(\vec{r}_2,0) |-\rangle _{\hat{b}}$ respectively. Thus the \textit{preparation-determined (epistemic) quantum state plays a purely statistical role in our model}. We will see later that the statistical relationship between the epistemic and ontic quantum states is preserved with time (see equations \ref{d1} and \ref{d2}). \\

Our second assumption is about the initial distribution of the positions of particles. Consider an ensemble of pairs of particles having the same joint ontic quantum state $\chi_1(\vec{r}_1,0) |i_1\rangle_{\hat{a}}\otimes\chi_2(\vec{r}_2,0)|i_2\rangle_{\hat{b}}$. Let the initial distribution of positions for this ensemble be denoted by $\rho_{i_1i_2}(\vec{r}_1,\vec{r}_2,0)$. We assume that $\rho_{i_1i_2}(\vec{r}_1,\vec{r}_2,0)=|\chi_1(\vec{r}_1,0)|^2|\chi_2(\vec{r}_2,0)|^2$. Since $|\psi_o(t)\rangle$ evolves via the Schrodinger equation, the corresponding continuity equation\footnote{The Schrodinger equation $-\frac{\nabla^2\psi}{2m} +V(\vec{x})\psi = i\frac{\partial \psi}{\partial t}$ implies the continuity equation $\vec{\nabla} \cdot({|\psi|^2 \frac{\vec{\nabla} S}{m}}) + \frac{\partial |\psi|^2}{\partial t} =0$ where $\psi(\vec{x},t) =R(\vec{x},t)e^{iS(\vec{x},t)}$. Here $\vec{x}$ represents a point in, and $\vec{\nabla}$ is acting on, the configuration space.} defines time evolution of the respective ensemble distribution $\rho_{i_1i_2}(\vec{r}_1,\vec{r}_2,t)$. The distribution of positions over all the ensembles at any time is given by
\begin{align}
\rho(\vec{r}_1,\vec{r}_2,t) = \sum_{i_1,i_2} |c_{i_1i_2}|^2 \rho_{i_1i_2}(\vec{r}_1,\vec{r}_2,t) \label{d5}
\end{align}

Note that: \textit{a)} the distribution of the joint ontic quantum states given by equation \ref{d3} is identical to the distribution of ($\lambda'_i$, $\lambda'_j$) in equation \ref{o}; and \textit{b)} the spin eigenket $|i\rangle$ of the future measurement setting, contained in the ontic quantum state, determines the local measurement outcome analogous to the hidden variable $\lambda'_i$ in equation \ref{a'}. This establishes the connection to the Brans model, which was originally proposed as superdeterministic. \\


Now let us describe the measurement process. First, the measuring apparatus creates a correlation between the positions of particles and their spins (along the directions chosen by experimenters). For this stage of the measuring process, we assume an interaction Hamiltonian $\hat{H}_I= g(\hat{p}_{\hat{x}_1}\otimes\hat{\sigma}_{\hat{a}}\otimes\hat{I}\otimes\hat{I} + \hat{I}\otimes\hat{I}\otimes\hat{p}_{\hat{x}_2}\otimes\hat{\sigma}_{\hat{b}})$. Here \textit{g} is a constant proportional to the strength of interaction, and $\hat{p}_{\hat{x}_1}$ and $\hat{p}_{\hat{x}_2}$ are the momenta conjugate to $\hat{x}_1$ and $\hat{x}_2$ respectively\footnote{Here $\vec{r}_1\equiv x_1 \hat{x} + y_1\hat{y} + z_1 \hat{z}$ and $\vec{r}_2 \equiv x_2 \hat{x} + y_2\hat{y} +z_2 \hat{z}$, where $\hat{x},\hat{y},\hat{z}$ are unit vectors along $x$, $y$ and $z$ axes respectively.}. The constant \textit{g} is assumed to be large enough so that, in the time interval $\hat{H}_I$ is acting, the remaining terms in the Hamiltonian can be ignored, i.e $\hat{H} \approx \hat{H}_I$. Let us first consider the evolution of the epistemic quantum state $|\psi_e(0)\rangle$:
\begin{align}
&\langle \vec{r}_1| \langle \vec{r}_2 |\psi_e(t)\rangle = \langle \vec{r}_1| \langle \vec{r}_2 | e^{-i\hat{H}t} |\psi_e(0)\rangle\\
& = c_{++} \chi_1(\vec{r}_1 -gt\hat{x},0) \chi_2(\vec{r}_2-gt\hat{x},0)|+\rangle_{\hat{a}} |+\rangle _{\hat{b}} + c_{+-} \chi_1(\vec{r}_1 -gt\hat{x},0) \chi_2(\vec{r}_2+gt\hat{x},0)|+\rangle_{\hat{a}}|-\rangle _{\hat{b}}  \nonumber\\
&+c_{-+} \chi_1(\vec{r}_1 +gt\hat{x},0) \chi_2(\vec{r}_2-gt\hat{x},0)|-\rangle_{\hat{a}}|+\rangle _{\hat{b}} + c_{--} \chi_1(\vec{r}_1 +gt\hat{x},0) \chi_2(\vec{r}_2+gt\hat{x},0)|-\rangle_{\hat{a}}|-\rangle _{\hat{b}} \label{d1}
\end{align}
We see from the above expression that over time the configuration space wavepacket evolves into four effectively disjoint eigenpackets. \\

Now consider what happens to an ontic quantum state $|\psi_o(0)\rangle = |\chi_1\rangle|i_1\rangle_{\hat{a}}\otimes |\chi_2\rangle|i_2\rangle_{\hat{b}}$. From the Schrodinger equation, using the same interaction Hamiltonian we find
\begin{align}
\langle \vec{r}_1| \langle \vec{r}_2 |\psi_o(t)\rangle &= \langle \vec{r}_1| \langle \vec{r}_2| e^{-i\hat{H}t} |\psi_o(0)\rangle\\
&= \chi_1(\vec{r}_1 -i_1gt\hat{x},0)|i_1\rangle_{\hat{a}} \otimes\chi_2(\vec{r}_2 -i_2gt\hat{x},0)|i_2\rangle_{\hat{b}}\label{d2}
\end{align}

We see that the joint ontic quantum state remains factorisable at all times, and that the single-particle wavepackets $\chi_1(\vec{r}_1, t)$ and $\chi_2(\vec{r}_2, t)$ separate in physical space in a manner that depends on the ontic spin states $|i_1\rangle_{\hat{a}}$ and $|i_2\rangle_{\hat{b}}$ respectively. Further, since the single-particle wavepackets act as pilot waves for the corresponding particles (from equation \ref{d4}), the particle trajectories also separate in physical space. From equations \ref{d1} and \ref{d2}, we note that, as expected, $|\psi_e(t)\rangle$ continues to describe an ensemble distribution of various $|\psi_o(t)\rangle$'s over time. The ensemble distribution $|c_{i_1i_2}|^2$ of ontic quantum states remains constant throughout since $|\langle \psi_o(0)|\psi_e(0)\rangle|^2 = |\langle \psi_o(t)|\psi_e(t)\rangle|^2$.\\

After the wavepackets corresponding to different spin eigenvalues have sufficiently separated from each other, the positions of the particles are measured. This is usually in the form of a photographic plate on which the particles impinge after the interaction Hamiltonian has been turned off. Since, for a particular joint ontic quantum state, the distribution of positions is given by $\rho_{i_1i_2}(\vec{r}_1,\vec{r}_2,t)=|\chi_1(\vec{r}_1 -i_1gt\hat{x},0)|^2|\chi_2(\vec{r}_2 -i_2gt\hat{x},0)|^2$, each particle impinges on the plate in the appropriate region, which allows us to discern which wavepacket it belonged to and hence its spin. The probability of obtaining a particular pair of results $\{i_1,i_2\}$ is equal to the probability of having a particular joint ontic quantum state in the ensemble of pairs of particles. The latter probability equals $|c_{i_1i_2}|^2$, and the Bell correlations are thus exactly reproduced.\\

\section{Effective nonlocal signalling in non-equilibrium}
The discussion up till now has assumed a particular initial distribution of hidden variables that exactly reproduces the Bell correlations. We now discuss a `non-equilibrium' extension of our model having an arbitrary distribution of hidden variables. The dynamics of the model is kept unchanged: the joint ontic quantum state and the epistemic quantum state evolve via the Schrodinger equation, and the position of each particle is guided locally by its corresponding 3-space wavepacket just as before.\\

Our model has two distinct hidden-variable distributions - the distribution of positions of particles in 3-space, and the distribution of ontic quantum states. We separately consider non-equilibrium for these two distributions. 
\subsection{Non-equilibrium for the distribution of positions}
Suppose the initial distributions of positions are given by arbitrary $\rho_{i_1i_2}(\vec{r}_1,\vec{r}_2,0)$, $i_1,i_2 \in \{+,-\}$, instead of $ |\chi_1(\vec{r}_1,0)|^2 |\chi_2(\vec{r}_2,0)|^2$, while the distribution of ontic quantum states remains in equilibrium. The position distributions evolve via the equation 
\begin{align}
\frac{\partial \rho_{i_1i_2}(\vec{r}_1,\vec{r}_2,t)}{\partial t} + \vec{\nabla} \cdot({\rho_{i_1i_2}(\vec{r}_1,\vec{r}_2,t) \frac{\vec{\nabla} S_{i_1i_2}}{m}})  =0
\end{align}
where the density $|\chi_1(\vec{r}_1-i_1gt\hat{x},0)|^2 |\chi_2(\vec{r}_2-i_2gt\hat{x},0)|^2$ has been replaced by $\rho_{i_1i_2}(\vec{r}_1,\vec{r}_2,t)$ in the continuity equation. It is clear that, as long as the interaction Hamiltonian acts for a sufficient period of time, the trajectories of particles belonging to different ontic quantum states will separate, regardless of the initial distribution of position. Thus the final positions where the particles strike the photographic plate will continue to yield unambiguous measurement results. Since the distribution of measurement outcomes is fixed by the distribution of ontic quantum states, the outcome probabilities remain unchanged. Hence a violation of no-signalling predicated on outcome probabilities is ruled out. However, as we show below, the local (marginal) position distribution, which determines the \textit{shapes} of spots formed on the photographic plate over time at one wing, will depend on the measurement setting at the other wing. Thus, no-signalling predicated on position probabilities will still be violated\footnote{In general, no-signalling is violated if the local probability of an event $A$ depends non-trivially on an event $B$ which is space-like separated from the event $A$, i.e  $p(A|B) \neq p(A|B')$.}$^,$\footnote{Since measurement outcomes are inferred from position measurements, it is logically impossible to have signalling in the outcome distribution without signalling in the position distribution. If there is signalling in the position distribution without signalling in the outcome distribution, only the shapes of spots at one wing can have a non-trivial dependence on the measurement setting at the other wing. }.\\

Consider for instance the shapes of spots on the photographic plate corresponding to the local outcomes $|+\rangle_{\hat{a}}$ and $|-\rangle_{\hat{a}}$. These will be determined by the local distribution of position of the first particle over all ensembles. From equation \ref{d5} 
\begin{align}
\rho(\vec{r}_1,t) &\equiv \int d\vec{r}_2\textbf{ } \rho(\vec{r}_1,\vec{r}_2,t) = \sum_{i_1,i_2} |c_{i_1i_2}|^2 \int d\vec{r}_2 \textbf{ } \rho_{i_1i_2}(\vec{r}_1,\vec{r}_2,t) \equiv \sum_{i_1,i_2} |c_{i_1i_2}|^2 \rho_{i_1i_2}(\vec{r}_1, t)\label{e}
\end{align}
For the singlet state, we know that 
\begin{equation}
\begin{aligned}
&|c_{++}|^2= |c_{--}|^2=\frac{1-\hat{a}\cdot\hat{b}}{4}\\
&|c_{+-}|^2= |c_{-+}|^2=\frac{1+\hat{a}\cdot\hat{b}}{4}\label{hagga}
\end{aligned}
\end{equation}
Plugging in the values in equation \ref{e}, we find

\begin{align}
\rho(\vec{r}_1,t)  = \frac{1-\hat{a}\cdot\hat{b}}{4}\rho_{++}(\vec{r}_1,t) + \frac{1+\hat{a}\cdot\hat{b}}4\rho_{+-}(\vec{r}_1,t) +\frac{1+\hat{a}\cdot\hat{b}}{4}\rho_{-+}(\vec{r}_1,t) + \frac{1-\hat{a}\cdot\hat{b}}{4}\rho_{--}(\vec{r}_1,t) \label{f}
\end{align}
We know that, in the case of the equilibrium distribution
\begin{align}
\rho(\vec{r}_1,t) = \frac{|\chi_1(\vec{r}_1-gt\hat{x},0)|^2+|\chi_1(\vec{r}_1+gt\hat{x},0)|^2}{2}
\end{align}
so that the shape of the spot corresponding to $|+\rangle_{\hat{a}}$ is given by $|\chi_1(\vec{r}_1-gt\hat{x},0)|^2$, and that corresponding to $|-\rangle_{\hat{a}}$ is given by $|\chi_1(\vec{r}_1+gt\hat{x},0)|^2$. Both the shapes are independent of the measurement settings. But if $\rho_{i_1i_2}(\vec{r}_1,t)$ are arbitrary, then it is clear from equation \ref{f} that the local position distribution $\rho(\vec{r}_1,t)$ will depend on the measurement setting chosen at the other wing of the experiment. Given that the outcome distribution has no such dependence, we conclude that the \textit{shapes of spots} formed on the photographic plate will be influenced by the measurements setting at the other wing. This will constitute a signal from one wing of the experiment to the other.\\

\subsection{Non-equilibrium for the distribution of ontic quantum states}
Let us now consider the case of a non-equilibrium distribution of only the ontic quantum states. The equilibrium distribution is given by the modulus squared of the coefficients $c_{i_1i_2}$, $i_1, i_2 \in \{+,-\}$, in equation \ref{d3}. Consider a non-equilibrium distribution defined by a different set of coefficients $c'_{i_1i_2}$ having the following relationship with the equilibrium distribution
\begin{equation}
\begin{aligned}
&|c'_{++}|^2 = |c_{++}|^2 + \frac{|c_{--}|^2}{3} \\
&|c'_{+-}|^2 = |c_{+-}|^2 +\frac{|c_{--}|^2}{3}\\
&|c'_{-+}|^2 = |c_{-+}|^2 \\
&|c'_{--}|^2 = \frac{|c_{--}|^2}{3} \label {haggu}
\end{aligned}
\end{equation}
Since, as noted in the previous section, the equilibrium distribution $|c_{i_1i_2}|^2$ is time-independent, the non-equilibrium distribution $|c'_{i_1i_2}|^2$ as defined above is also time-independent\footnote{ We do not concern ourselves here with the question of a relaxation mechanism to the equilibrium distribution. We also note that perhaps ‘equilibrium’ might not be the best term for many retrocausal models, because it incorporates a notion of an arrow of time in the word itself. For a complete discussion of quantum equilibrium, please refer to the references given in the Introduction.}. Now consider the local probability of getting a $|+\rangle_{\hat{a}}$ outcome. This will be equal to $|c'_{++}|^2 + |c'_{+-}|^2 = |c_{++}|^2 +|c_{+-}|^2 +2\times \frac{|c_{--}|^2}{3}$. Using equation \ref{hagga}, this turns out to be $\frac{4-\hat{a}\cdot\hat{b}}{6}$. The expression depends on the measurement setting at the other wing $\hat{b}$, violating the no-signalling constraints predicated on outcome probabilities. \\
\begin{figure}
\includegraphics[scale=0.7]{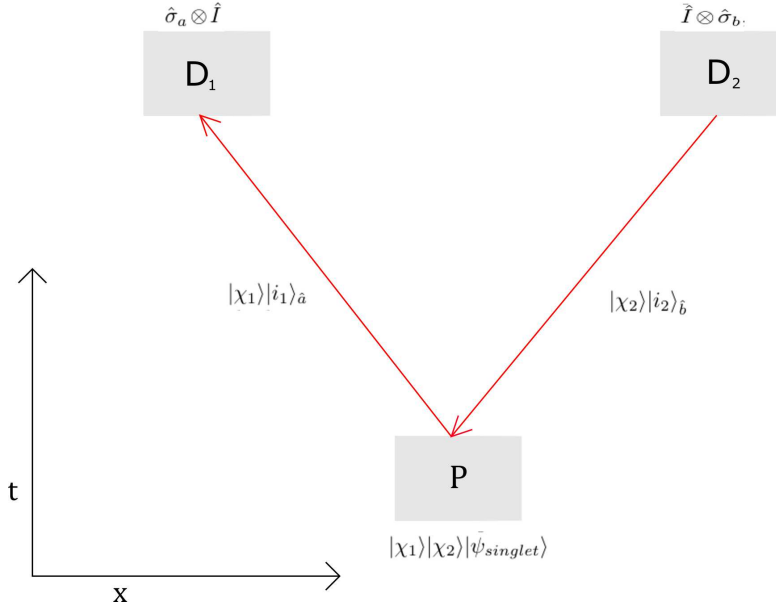}
\caption{Schematic illustration of `effective nonlocal signalling' in non-equilibrium. The red lines in the figure indicate the flow of information from $D_2$ to $D_1$. The measurement setting $\hat{I}\otimes\hat{\sigma}_{\hat{b}}$ made in the space-time region indicated by $D_2$ retrocausally influences the distribution of hidden variables at preparation source P. This distribution in turn influences the local probabilities at $D_1$ (in non-equilibrium). Though the signal is nonlocal, the underlying dynamics is local.}
\end{figure}

Will the shapes of spots formed on the photographic plate at one wing also depend on the measurement setting at other wing? Replacing  $|c_{i_1i_2}|^2$ by  $|c'_{i_1i_2}|^2$ in equation \ref{d5} and using equations \ref{hagga} and \ref{haggu}, the marginal distribution of the position of the first particle turns out to be 
\begin{align}
\rho(\vec{r}_1,t) = \frac{4-\hat{a}\cdot\hat{b}}{6}|\chi_1(\vec{r}_1-gt\hat{x})|^2 + \frac{2+\hat{a}\cdot\hat{b}}{6}|\chi_1(\vec{r}_1+gt\hat{x})|^2
\end{align}
which indicates that the shape of the spot corresponding to $|+\rangle_{\hat{a}}$ is given by $|\chi_1(\vec{r}_1-gt\hat{x},0)|^2$, while that corresponding to $|-\rangle_{\hat{a}}$ is given by $|\chi_1(\vec{r}_1+gt\hat{x},0)|^2$. Both the shapes are independent of the measurement settings (only the relative proportion of outcomes depends on the measurement settings). Thus, in the case of a non-equilibrium distribution of the ontic quantum states, there is no effect on the shapes of spots formed on the photographic plate.\\

The nonlocal transfer of information, in either case of non-equilibrium, is achieved by a Lorentz-covariant local dynamics. 
The measurement setting $\hat{\sigma}_{\hat{b}}$ retrocausally influences the distribution of positions (ontic quantum states) at the time of preparation, and this in turn influences the local position probabilities (local outcome probabilities) at the other wing, at a space-like separated point, via a `zigzag' path in space-time not exceeding the speed of light (see Fig. 2). Since the local probabilities depend on an event that is space-like separated, we may term it as `effective nonlocal signalling'.

\section{Discussion and Conclusion}
Each particle in our model has an ontology consisting of position in 3-space and an ontic quantum state. It might at first appear that position is not necessary as a hidden variable, since the ontic quantum state already has a spin eigenket which determines the measurement outcome. But without including position in the ontology, there would be no way to account for the final spot on the photographic plate without a collapse of the 3-space wavepacket (in this model).\\

It might also be mistakenly thought that the model is $\psi$-ontic since there is an ontic quantum state $|\psi_o(t)\rangle$ in the hidden-variable description. But this state must be distinguished from the preparation-determined (epistemic) quantum state $|\psi_e(t)\rangle$. The set of possible $|\psi_o(t)\rangle$'s in an experiment is determined only by the future measurement settings. It is only in the ensemble distribution of different $|\psi_o(t)\rangle$'s that $|\psi_e(t)\rangle$ plays a role in our model. This can be readily seen if we prepare two different epistemic quantum states (say a singlet state $|\psi_e(0)\rangle_{singlet}$ and a triplet state $|\psi_e(0)\rangle_{triplet}$) and subject both to the same Bell measurement. The set of possible $|\psi_o(0)\rangle's$ will be identical, reflecting overlap in the hidden-variable space of $|\psi_e(0)\rangle_{singlet}$ and $|\psi_e(0)\rangle_{triplet}$. In other words, given knowledge of the hidden variable $|\psi_o(0)\rangle$, it will be impossible to determine which preparation-determined quantum state it belongs to. Thus our model is by definition $\psi$-epistemic \cite{harrikens, pbr}. Further, our ontic quantum state is always factorisable and contains 3-space wavepackets for the two particles, whereas the preparation-determined quantum state is entangled and contains a configuration space wavepacket in general.\\


We have discussed the signalling properties of our model given a non-equilibrium distribution of the hidden variables. If only the distribution of the positions of particles is in non-equilibrium, the local position probabilities at one wing depend on the measurement setting at the other wing, but the local outcome probabilities are unaffected. This leads to the following effect: the shapes of spots formed on the photographic plate at one wing are influenced by the measurement setting at the other wing. If instead, only the distribution of the ontic quantum states is in non-equilibrium, the local outcome probabilities at one wing depend on the measurement setting at the other wing, but the shapes of the spots are unaffected. Hence, non-equilibrium in each hidden variable distribution causes no-signalling to be violated in a different way. Since the dynamics of the model is local throughout, we conclude that retrocausality may provide a means for such violations while retaining Lorentz covariance at the hidden variable level. From our viewpoint this is an attractive positive feature of retrocausal hidden-variable models which suggests a solution to the problem of fine-tuning pointed out by Wood and Spekkens\cite{woodspek}. Unlike other authors \cite{almada, prituning} who have appealed to the symmetries in retrocausal models in order to justify the fine-tuning, we believe that a more straightforward answer can be given by rejecting fine-tuning as an inevitable feature of retrocausal models and no-signalling as a fundamental feature of Nature. Then the task ahead would be two-fold. First, to give an explanation why the quantum systems accessible to us have the equilibrium no-signalling distribution of hidden variables instead of an arbitrary signalling distribution. Such an explanation can be either dynamical, in which case the emergence over time of the no-signalling equilibrium distribution from an arbitrary signalling distribution will have to be shown, or it can be all-at-once\cite{whartonmain}, in which case the emergence of no-signalling equilibrium distribution will have to be shown as a consequence of boundary conditions both in the past and the future. Second, to address the apparent paradoxes involving retrocausal signalling possible for a non-equilibrium distribution.\\

Our model has a connection to the modal class of interpretations of QM \cite{sep}. These describe a quantum system by two states, a \textit{dynamical state} and a \textit{value state}. The dynamical state determines which physical properties the quantum system may possess, while the value state determines which physical properties the system actually possesses, at a certain instant. The dynamical state is identified as the usual quantum state in Hilbert space, but the definition of the value state depends on the particular modal interpretation. The dynamical state evolves via the Schrodinger equation, while the value state usually has a more complicated evolution law. We see that our model fits into the category of a `modal interpretation of Bell correlations' but with retrocausality. The state $|\psi_e(t)\rangle$ is the dynamical state as in other modal interpretations. We identify $\big (\vec{r}_1(t),\vec{r}_2(t), |\psi_o(t)\rangle\big )$ as the value state for our model. Analogous to modal interpretations, our value state determines which physical properties are actually possessed by the quantum system of two spin-1/2 particles subjected to Bell measurements. These are the positions of particles in 3-space $\vec{r}_1(t)$ and $\vec{r}_2(t)$, the 3-space ontic wavepackets $\chi_1(\vec{r}_1,t)$ and $\chi_2(\vec{r}_2,t)$, and the ontic spin eigenstates of the future measurement settings $|i_1\rangle_{\hat{a}}$ and $|i_2\rangle_{\hat{b}}$. The positions evolve via equation \ref{d4}, whereas the joint ontic quantum state evolves via the Schrodinger equation. The dynamical state determines the probability of a particular value state via equations \ref{d3} and \ref{d5} (in equilibrium). It can be asked if, like modal interpretations, our model treats the measurement process as an ordinary physical interaction. This can be answered only if the retrocausal mechanism alluded to in section \ref{main}, by which information about future measurement settings is made available retrocausally at the preparation source, is defined in physical terms rather than assumed in an \textit{ad hoc} manner as done presently. \\

The present work can be compared to some previous attempts in the literature to introduce wavefunctions in physical space. Sutherland \cite{sutherland} developed a local retrocausal de Broglie-Bohm type model which, under some conditions on the configuration space wavefunction at a future time, assigns a 3-space wavefunction to each particle in the past. However, the probability density for position is not non-negative in that model. Norsen et al. \cite{norsen10, norsen15} use conditional de Broglie-Bohm wavefunctions, which can be argued to represent wavefunctions in physical space, to develop two models for spinless particles. One of these requires a highly redundant ontic space in order to reproduce QM predictions. The other has a reduced ontological complexity at the cost of reproducing QM predictions only approximately. Both models have nonlocal interactions between the particles. Gondran et al. \cite{gondran} develop a nonlocal model for Bell correlations which attributes a 3-space wavefunction to  each particle, but the full quantum state is part of the hidden-variable description (hence the model is $\psi$-ontic). In contrast, the model we have presented does not suffer from negative probabilities, exactly reproduces the QM predictions without a high ontological complexity, is local as regards ontology and interactions between particles, and has a clean ontological separation between the single particle ontic quantum states with 3-space wavepackets and the epistemic preparation-determined quantum state with configuration space wavepackets\footnote{It is interesting to note that the prototype of our model was pre-empted by Corry \cite{prempt}, who however did not formally develop the idea.}. However, the model as currently presented is restricted to Bell correlations, and the retrocausality is assumed in an \textit{ad hoc} manner.\\

\begin{acknowledgements}
I would like to thank my thesis advisor Antony Valentini for stimulating discussions and helpful suggestions throughout this work, and for encouragement to work on retrocausality. I would also like to thank Ken Wharton and an anonymous referee for several helpful comments and discussions on an earlier draft of the paper. 
\end{acknowledgements}
\bibliographystyle{bhak}
\bibliography{bib}


\end{document}